\documentclass[a4paper,preprint,12pt,nofootinbib]{revtex4}

\usepackage[tbtags]{amsmath}
\usepackage{amssymb}
\usepackage{amsthm}
\usepackage{fancyhdr}
\usepackage[dvips]{graphicx}
\usepackage{color}
\usepackage{exscale}

\newcommand{\intd}{\int [dk]\,}
\newcommand{\intdtwo}{\int [dk][dl]\,}
\newcommand{\Cl}{{\text{Cl}_2}}
\newcommand{\Clthree}{{\text{Cl}_3}}
\newcommand{\Li}{\text{Li}_2}
\newcommand{\Lithree}{\text{Li}_3}

\newcommand{\Heff}{\mathcal{H}_\text{eff}}
\newcommand{\Op}{O}

\allowdisplaybreaks[1]

\begin{document}
\title{Analytic calculation of two-loop QCD corrections to 
$b\to s\ell^+\ell^-$ in the high $q^2$ region}
\author{C. Greub} 
\email{christoph.greub@itp.unibe.ch}
\author{V. Pilipp}
\email{volker.pilipp@itp.unibe.ch}
\author{C. Sch\"upbach}
\email{christof.schuepbach@itp.unibe.ch}
\affiliation{Center for Research and Education in Fundamental Physics,
             Institute for Theoretical Physics, University of Bern\\
             Sidlerstrasse 5, CH-3012 Bern }

\begin{abstract}
We present our results for the  NNLL virtual corrections to the matrix 
elements of the operators $\Op_1$ and
$\Op_2$ for the inclusive process $b\to s\ell^+\ell^-$ in the kinematical
region $q^2>4m_c^2$, where $q^2$ is the invariant mass squared of the 
lepton-pair. 
This is the first analytic two-loop calculation of these matrix
elements in the high $q^2$ region. 
We give the matrix elements as an expansion in $m_c/m_b$ and keep the
full analytic dependence on $q^2$.  Making extensive use of
differential equation techniques, we fully automatize the expanding of
the Feynman integrals in $m_c/m_b$. 
In coincidence with an earlier work where the master integrals were
obtained numerically \cite{Ghinculov:2003qd}, 
we find that in the high $q^2$ region the $\alpha_s$ corrections
to the matrix elements $\langle s \ell^+ \ell^-|O_{1,2}|b\rangle$ 
calculated in the
present paper lead to a decrease of the perturbative part of the
$q^2$-spectrum by $10\%-15\%$
relative to the NNLL result in which these contributions are put to zero
and reduce the renormalization scale uncertainty to $\sim 2\%$.
\end{abstract}

\maketitle
\section{Introduction}
Flavor-changing neutral currents play an important role in the
indirect search for new physics. For inclusive decays there exists the
framework of operator-product expansion, which makes theoretically
clean predictions possible. Of special interest in this context is the 
decay mode $B\to
X_s\ell^+\ell^-$. In the regions where the lepton invariant mass squared
$q^2$ is far away from the $c\bar{c}$-resonances, 
the dilepton invariant mass spectrum and the forward-backward
asymmetry can be precisely predicted.

The status of the calculation of these observables 
is the following: The leading logarithmic (LL) and the
next-to-leading logarithmic (NLL) QCD contributions were calculated in 
\cite{Grinstein:1988me,Misiak:1992bc,Buras:1994dj}. Next-to-next-to-leading
logarithmic (NNLL) corrections to the Wilson coefficients at the
matching scale $\mu \sim m_W$, which required to perform  
two-loop matching calculations of the full standard model (SM) theory
onto the effective theory, have been worked out in
\cite{Buchalla:1995vs,Adel:1993ah,Greub:1997hf,Bobeth:1999mk}.
The anomalous dimensions matrices needed to obtain the Wilson coefficients
at the low scale $\mu \sim m_b$ (requiring up to three-loop calculations
for certain entries) were obtained in 
\cite{Buchalla:1995vs,Chetyrkin:1996vx,Gambino:2003zm,Bobeth:2003at,
Gorbahn:2004my}.
NNLL QCD corrections at the level of the matrix elements of the
operators involved were calculated for
the dilepton invariant mass spectrum and for the forward-backward asymmetry
in \cite{Asatrian:2001de,Asatryan:2001zw,Asatryan:2002iy,Ghinculov:2002pe,
Asatrian:2002va,Ghinculov:2003qd}.
Power corrections of the order $1/m_b^2$, $1/m_c^2$ and $1/m_b^3$ 
have been worked out in 
\cite{Falk:1993dh,Ali:1996bm,Chen:1997dj,Buchalla:1997ky,Buchalla:1998mt,
Bauer:1999kf,Ligeti:2007sn}.
Finally, in \cite{Huber:2005ig,Huber:2007vv,Huber:2008ak} certain 
classes of logarithmically
enhanced electromagnetic corrections were taken into account.
 
So far, {\itshape analytic results} for the NNLL QCD corrections to the matrix 
elements associated with
the operators $\Op_1$ and $\Op_2$ are only available in the region of low $q^2$.
The corresponding results were obtained as a double-expansion 
in $m_c/m_b$ and $q^2/m_b^2$ 
\cite{Asatrian:2001de,Asatryan:2001zw,Asatryan:2002iy}.
The present paper deals with the NNLL QCD corrections in the high 
$q^2$ region, i.e.
$q^2>4m_c^2$. In particular we evaluate {\itshape virtual} QCD corrections to 
the matrix elements of the
operators $\Op_1$ and $\Op_2$ at order $\alpha_s$. In contrast to 
\cite{Ghinculov:2003qd}, where the relevant master integrals were
calculated numerically, we present these matrix elements 
as analytic functions of $m_c$ and $q^2$. The purpose of the present
paper is twofold: First, to deliver a non-trivial independent check
of the results found in \cite{Ghinculov:2003qd} and second, 
to provide the user with analytic
formulas in which the parameters ($m_c/m_b$ and $\mu/m_b$) 
and $q^2$ can easily be changed.
 
To get these analytic results, we perform an expansion in $m_c/m_b$ 
and keep the full 
analytic dependence on $q^2$. We expand the two-loop Feynman integrals
by combining method of regions
\cite{smirnov:2002pj,gorishnii:1989dd,beneke:1997zp,smirnov:1990rz} 
and differential equation techniques
\cite{remiddi:1997ny,pilipp:2007mg,Pilipp:2008ef,boughezal:2007ny}.
We end up with an expansion of $\langle s \ell^+ \ell^- |\Op_{1,2}|b\rangle$ up
to the 20th power in $m_c/m_b$. As the resulting expressions for
these matrix elements are rather lengthy, we are not able to print
them in the paper. We provide Mathematica and  c++ code of our results
in the source files of the present paper at arXiv. 

The well-known breakdown of the $\Lambda/m_b$ expansion at the
endpoint $q^2=m_B^2$ seems to question the relevance of the
perturbative contributions in the high $q^2$-region calculated 
in this paper. However, as it was shown in \cite{Bauer:2000xf} and 
\cite{Neubert:2000ch}
(illustated there for the analogous lepton invariant mass spectrum in the
inclusive semileptonic decay $B \to X_u \ell \nu$) 
that the integrated high $q^2$-spectrum allows for a modified version
of the heavy-quark expansion (the so-called hybrid expansion), our
present work is well-motivated.

The paper is organized as follows. Sections \ref{s1} and \ref{s2} are
dedicated to the technical details of the calculation. We give all
necessary definitions in Section \ref{s1}. In Section \ref{s2} we
explain the evaluation of the Feynman integrals in detail. In Section 
\ref{secres} we investigate the (numerical) stability of the expansion
in $m_c/m_b$, concluding that retaining terms up to the 20th power in 
$m_c/m_b$ leads to precise results. In this Section
we also discuss the numerical impact of our 
calculation on the
dilepton invariant mass spectrum. 
In coincidence with 
\cite{Ghinculov:2003qd} 
we find that in the high $q^2$ region the order $\alpha_s$ corrections
to the matrix elements $\langle s \ell^+ \ell^-|O_{1,2}|b\rangle$ 
calculated in the
present paper
lead to a decrease of the perturbative part of the
$q^2$-spectrum by $10\%-15\%$
relative to the NNLL result in which these corrections are put to zero
and reduce the renormalization scale uncertainty to $\sim 2\%$.

\section{Definitions \label{s1}}
\begin{figure}
\begin{center}
\resizebox{\textwidth}{!}{\includegraphics{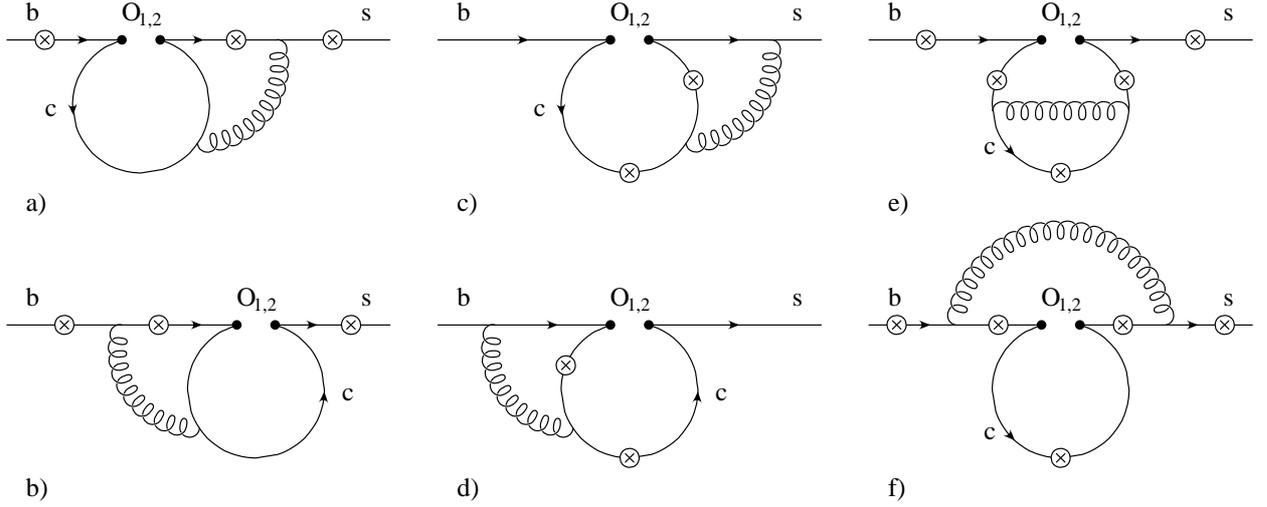}}
\end{center}
\caption{Diagrams that have to be taken into account 
  at order $\alpha_s$. The circle-crosses denote the possible
  locations where the virtual photon is emitted (see text).}
\label{f1}
\end{figure}

As in the previous paper \cite{Asatryan:2001zw} we write the
effective Hamiltonian that contributes to $B \to X_s \ell^+\ell^-$ in the
form 
\begin{equation}
\Heff=-\frac{4G_F}{\sqrt{2}}V_{ts}^*V_{tb}\sum_{i=1}^{10}C_i(\mu)\Op_i(\mu),
\label{heff}
\end{equation}
where we have neglected the small CKM combination $V^*_{us}V_{ub}$. 
The operator basis is defined as
\begin{equation} \label{oper}
\begin{array}{rclrcl}
    \Op_1    & = & (\bar{s}_L \gamma_\mu T^a c_L) (\bar{c}_L \gamma^\mu T^a b_L)
\, ,&
    \vspace{0.4cm}
    \Op_2    & = & (\bar{s}_L \gamma_\mu c_L) (\bar{c}_L \gamma^\mu b_L) \, ,\\
    \vspace{0.2cm}
    \Op_3    & = &  (\bar{s}_L \gamma_{\mu}  b_{L }) \sum_q (\bar{q}\gamma^{\mu}
q) \, ,&
    \vspace{0.2cm}
    \Op_4    & = &  (\bar{s}_{L}\gamma_{\mu} T^a b_{L }) \sum_q
(\bar{q}\gamma^{\mu} T^a q) \, ,\\
    \vspace{0.2cm}
    \Op_5    & = & (\bar{s}_L\gamma_\mu \gamma_\nu \gamma_\rho b_L)
                    \sum_q(\bar{q} \gamma^\mu \gamma^\nu \gamma^\rho q) \, ,&
    \vspace{0.2cm}
    \Op_6   & = & (\bar{s}_L\gamma_\mu \gamma_\nu \gamma_\rho T^a b_L)
                    \sum_q(\bar{q} \gamma^\mu \gamma^\nu \gamma^\rho T^a q) \,
,\\
    \vspace{0.2cm}
    \Op_7    & = & \frac{e}{g_s^2} m_b (\bar{s}_{L} \sigma^{\mu\nu} b_{R})
F_{\mu\nu} \, ,&
    \vspace{0.2cm}
    \Op_8    & = & \frac{1}{g_s} m_b (\bar{s}_{L} \sigma^{\mu\nu} T^a b_{R})
G_{\mu\nu}^a \, ,\\
    \vspace{0.2cm}
    \Op_9    & = & \frac{e^2}{g_s^2}(\bar{s}_L\gamma_{\mu} b_L)
\sum_l(\bar{l}\gamma^{\mu}l) \, ,&
    \Op_{10} & = & \frac{e^2}{g_s^2}(\bar{s}_L\gamma_{\mu} b_L)
\sum_l(\bar{l}\gamma^{\mu} \gamma_{5}l) \, ,
\end{array}
\end{equation}
where the subscripts $L$ and $R$ refer to left- and right- handed components of
the fermion fields. The ingredients to obtain the Wilson coefficients 
$C_i$ at the scale $\mu$ of order $m_b$ can be found e.g. in 
\cite{Buchalla:1995vs,Bobeth:1999mk,Gambino:2003zm}.

In the present publication we calculate the virtual $\alpha_s$-corrections to
the matrix elements of $\Op_1$ and $\Op_2$ in the large $q^2$
region. 
Using equations of motion, we write these $\alpha_s$-corrections
in the form\footnote{Note that because of the extra factors $1/g_s^2$
  in the definition of $\Op_7$ and $\Op_9$ (\ref{formfact}) is indeed
  of order $\alpha_s$.}
\begin{equation}
\langle s\ell^+\ell^-|O_i|b\rangle_\text{2-loops} = 
-\left(\frac{\alpha_s}{4\pi}\right)^2
\left[
F_i^7\langle O_7\rangle_\text{tree}+
F_i^9\langle O_9\rangle_\text{tree}
\right].
\label{formfact}
\end{equation} 
The diagrams that contribute at order $\alpha_s$ to $b\to
s\ell^+\ell^-$ are shown in Fig.\ref{f1}.
By definition, we include in $F_{1,2}^{(7,9)}$ only
the contributions from the diagrams in Fig.~\ref{f1}a-e. As in
\cite{Asatryan:2001zw}, we absorb the contribution from Fig.~\ref{f1}f
into a modified Wilson coefficient $C_9$. This procedure is convenient,
because only the diagram Fig.~\ref{f1}f contains infrared
divergences. 

The ultraviolet renormalization works analogously to
\cite{Asatryan:2001zw}. In particular we use the same evanescent
operators. We use on-shell renormalization for the $s$- and the
$b$-field and renormalize $m_c$ in the pole mass scheme.

The kinematics is defined as follows:
We denote the momentum of the incoming $b$-quark by $p$ and the
momentum of the virtual photon by $q$. The momenta of the external
fermions are on-shell such
that $p^2=m_b^2$ and $(p-q)^2=0$, because we neglect the strange-quark
mass. Furthermore we use the notations
\begin{equation}
\hat{s}=\frac{q^2}{m_b^2}\quad\text{and}\quad
z=\frac{m_c^2}{m_b^2}.
\end{equation}

\section{Calculation of the master integrals
\label{s2}}
In the present section we explain for every diagram appearing in
Fig.~\ref{f1} the way we evaluated the master integrals 
that are specific to it. In Appendix \ref{cm} we list 
the master integrals that appear in more than one diagram and which
are straightforward to calculate.

We use the following notation 
\begin{equation}
[dk]=\left(\frac{\mu^2}{4\pi}e^{\gamma_E}\right)^\epsilon\frac{d^dk}{(2\pi)^d},
\label{mi0}
\end{equation}
where $d=4-2\epsilon$.
%We assume implicitly
%that every denominator contains a positive imaginary part $+i0$. 

For simplicity we set $m_b=1$ in the calculation of the master integrals, 
such that $m_c^2=z$. The dependence of the master integrals on
$m_b$ can be easily restored by dimensional analysis.

\subsection{General remarks about calculation techniques}
The Feynman integrals appearing in our calculation have been reduced
to a set of master integrals using the following methods:
Tensor integrals i.e.\ integrals containing Lorentz indices have been
reduced to scalar integrals via the Passarino-Veltman reduction scheme
\cite{Passarino:1978jh}. Finally these scalar integrals can be further
reduced by integration by parts (IBP) identities
\cite{Chetyrkin:1981qh,Tkachov:1981wb}. In particular we
used the algorithm described in \cite{Laporta:2001dd}. 
To this end we used the Maple implementation AIR
\cite{Anastasiou:2004vj} 
and a Mathematica implementation developed by us.
Since we consider the region $\hat{s}>4z$, we expanded the master
integrals in $z$ and kept the full analytic dependence in $\hat{s}$.

For power expanding Feynman integrals we use a combination of
\emph{method of regions} 
\cite{smirnov:2002pj,gorishnii:1989dd,beneke:1997zp,smirnov:1990rz} 
and \emph{differential equation techniques} 
\cite{Kotikov:1990kg,remiddi:1997ny,pilipp:2007mg,Pilipp:2008ef,
boughezal:2007ny}.
We consider a set of Feynman integrals $I_1,\ldots,I_n$ that depend on
the expansion parameter $z$ and that are related
by a system of differential equations:
\begin{equation}
\frac{d}{dz}I_\alpha = \sum_\beta h_{\alpha\beta} I_\beta + g_\alpha.
\label{mi1}
\end{equation}
We obtain (\ref{mi1}) by
differentiating $I_\alpha$ with respect to $z$ and applying IBP identities,
from where 
we obtain the original set of integrals and further integrals
contained in $g_\alpha$, which are simpler than $I_\alpha$ and have
been calculated before. Expanding the objects appearing in (\ref{mi1})
in $\epsilon$, $z$ and $\ln z$
\begin{eqnarray}
I_\alpha &=& \sum_{i,j,k} I_{\alpha,i}^{(j,k)} 
\epsilon^i z^j (\ln z)^k
\nonumber\\
h_{\alpha\beta}&=&\sum_{i,j} h_{\alpha\beta,i}^{(j)}
\epsilon^iz^j
\nonumber\\
g_\alpha&=&\sum_{i,j,k} g_{\alpha,i}^{(j,k)}
\epsilon^iz^j(\ln z)^k,
\label{mi2}
\end{eqnarray}
and inserting (\ref{mi2}) into (\ref{mi1}) we obtain algebraic
equations for the coefficients $I_{\alpha,i}^{(j,k)}$
\begin{equation}
  0=
  (j+1)I_{\alpha,i}^{(j+1,k)}+(k+1)I_{\alpha,i}^{(j+1,k+1)}-
  \sum_{\beta}\sum_{i^\prime}\sum_{j^\prime}
    h_{\alpha\beta,i^\prime}^{(j^\prime)}
    I_{\beta,i-i^\prime}^{(j-j^\prime,k)}
    -g_{\alpha,i}^{(j,k)}.
  \label{mi3}
\end{equation}
By means of (\ref{mi3}) we can reduce higher powers in $z$ of
$I_\alpha$ to lower powers. In practice this means that we need the
leading power and sometimes also the next-to-leading power of $I_\alpha$
as initial condition for (\ref{mi3}). We have
calculated these initial conditions by method of regions. 
Every region except the hard region leads to logarithms
in $z$. As we obtain the logarithms occurring at leading power
both from method of regions and from the recurrence relation (\ref{mi3}), 
differential equations provide a non
trivial check for method of regions,
i.e.\ we can make sure not to have forgotten or
counted twice any region.  

In (\ref{mi2}) we did not specify which values the summation index $j$
takes. Indeed we will have to deal with integrals that come with 
half-integer values of $j$ i.e.\ they have to be expanded in $\sqrt{z}$. 
On the other hand we have to presume that there exists $k_\text{max}$
such that $I^{(j,k)}_{\alpha,i}=0$ for all $k>k_\text{max}$ 
in order to solve (\ref{mi3}). 
We use the algorithm that was described in \cite{Pilipp:2008ef} to get the
possible values for $j$ and to determine $k_\text{max}$.
In addition this algorithm allows us to evaluate the
coefficients $I^{(j,k)}_{\alpha,i}$ numerically. We used this feature
to test the initial conditions.
 
In the following we will show in detail how to evaluate the master
integrals occurring from the diagrams in Fig.~\ref{f1} by this procedure.

\subsection{Diagrams of Fig. ~\ref{f1}a}
The topology of Fig.~\ref{f1}a contains in addition to (\ref{i22m}),
(\ref{i32z}), (\ref{i32z2}) and (\ref{i32z3}), which are easy to
evaluate, these two master integrals
\begin{eqnarray}
I_{a1} &=& \intdtwo\frac{1}{(k+p-q)^2(k+p)^2((k+l)^2-z)(l^2-z)}
\nonumber\\
I_{a2} &=& \intdtwo\frac{1}{(k+p-q)^2(k+p)^2((k+l)^2-z)(l^2-z)^2}
\label{mi4}
\end{eqnarray}  
where we use the notation (\ref{mi0}) and assume implicitly that every
denominator contains a positive imaginary part $+i0$.
We need both integrals in leading power i.e.\ at $z^0$. There are
three regions that contribute to this power:
The hard region $k^\mu,l^\mu\sim 1$, the soft region $k^\mu\sim 1$,
and $l^\mu\sim\sqrt{z}$ and the collinear region where both $k$ and
$l$ are collinear to $p-q$ (scaling see below).
Both integrals get a leading power
contribution in the hard region. The hard region corresponds to setting
$z=0$ in the integrand. In this limit we can reduce
$I_{a2}$ to $I_{a1}$ by IBP identities. 
$I_{a1}$ at $z=0$ can by evaluated via Feynman parameterization to
\begin{equation}
I_{a1,h} = -\frac{1}{(4\pi)^4}
\left(\mu^2e^{\gamma_E+i\pi}\right)^{2\epsilon} 
\frac{\Gamma(\epsilon)\Gamma(2\epsilon)
\Gamma^3(1-\epsilon)\Gamma(1-2\epsilon)}
{\Gamma(1+\epsilon)\Gamma(2-2\epsilon)\Gamma(2-3\epsilon)}
\,{}_2\text{F}_1(2\epsilon,1;1+\epsilon;1-\hat{s}),
\end{equation}
where
\begin{equation}
  {}_2\text{F}_1(a,b;c;x)=\frac{\Gamma(c)}{\Gamma(b)\Gamma(c-b)}
  \int_0^1dt\,t^{b-1}(1-t)^{c-b-1}(1-tx)^{-a}
\label{mi4.1}
\end{equation}
with $\Re c>\Re b>0$.
We used the Mathematica packages described in 
\cite{Huber:2005yg,Huber:2007dx} to obtain the expansion in $\epsilon$
of ${}_2\text{F}_1$.

In the soft region $k^\mu\sim 1$, $l^\mu\sim\sqrt{z}$ only $I_{a2}$
gets a leading power contribution:
\begin{equation}
I_{a2,s}=z^{-\epsilon}\intdtwo\frac{1}{(k+p-q)^2(k+p)^2k^2(l^2-1)^2}.
\label{mi5}
\end{equation}
Using IBP identities, (\ref{mi5}) can be reduced to a product of two
simple one-loop integrals.

Let us consider the collinear region. We introduce the following
light-like vectors $n_+$ and $n_-$, which fulfil $n_+^2=n_-^2=0$ and
$n_+\cdot n_-=1$. We define the decomposition of a Lorentz vector into
light-cone coordinates:
\begin{equation}
k^\mu=k^- n_-^\mu + k^+n_+^\mu+k_\perp^\mu
\end{equation}
where $k^{\pm}=k\cdot n_{\mp}$.
We choose $n_+$ to be collinear to $p-q$ and introduce the following
scaling
\begin{equation}
k^+,l^+\sim 1,\quad k_\perp,l_\perp\sim\sqrt{z}\quad\text{and}\quad
k^-,l^-\sim z.
\end{equation} 
As before, only $I_{a2}$ gets a leading power contribution
in this region. 
\begin{equation}
I_{a2,c}=z^{-2\epsilon}\intdtwo
\frac{1}
{(k+p-q)^2(2k^+p^-+1)((k+l)^2-1)(l^2-1)^2}.
\label{mi6}
\end{equation}
Via Feynman parameterization we evaluate (\ref{mi6}), obtaining
\begin{equation}
\frac{-1}{(4\pi)^4}\left(\mu^2e^{\gamma_E}\right)^{2\epsilon}
\frac{\Gamma^2(\epsilon)}{2\Gamma(1-\epsilon)}
\,{}_2\text{F}_1(1,1;2-\epsilon;1-\hat{s}).
\end{equation}
Finally the leading power contributions of the master integrals up to
order $\epsilon^0$ read
\begin{equation}
\begin{split}
&I_{a1}^{(0)}=\frac{1}{(4\pi)^4}\mu^{4\epsilon}
\Bigg[
-\frac{1}{2\epsilon^2}+
\frac{
\ln\left(\hat{s}\right)-i\pi-\frac{5}{2}
}{\epsilon}
\\
&\quad
-\frac{1}{2}\ln^2\left(\hat{s}\right)+(5+2i\pi)
   \ln\left(\hat{s}\right)+\text{Li}_2\left(1-\hat{s}\right)
   +\frac{13\pi^2}{12}-5i\pi-\frac{19}{2}
\Bigg]\\
&I_{a2}^{(0)}=\frac{1}{(4\pi)^4}\Bigg[
\frac{1}{2}\ln\left(\hat{s}\right)\ln^2(z)+
  \left(i
   \pi
   \ln\left(\hat{s}\right)+
    \text{Li}_2\left(1-\hat{s}\right)
   \right)\ln(z)
\\
&\quad
   -\frac{\pi^2}{2}\ln\left(\hat{s}\right)
   +i\pi 
   \text{Li}_2\left(1-\hat{s}\right)-
   \text{Li}_3\left(1-\hat{s}\right)
\Bigg].
\end{split}
\label{mi6.1}
\end{equation}

We continue with the calculation of the subleading powers of $I_{a1}$
and $I_{a2}$. By differentiating $I_{a1}$ and $I_{a2}$ with respect to $z$
and applying IBP identities we obtain a coupled system of differential
equations of the form (\ref{mi1}) with $h$ starting at order $\epsilon^0$
and $z^{-1}$. More explicitly (\ref{mi3}) becomes
\begin{equation}
  0=
  (j+1)I_{a\alpha,i}^{(j+1,k)}+(k+1)I_{a\alpha,i}^{(j+1,k+1)}-
  \sum_{\beta=1,2}\sum_{i^\prime=0}^{i+2}\sum_{j^\prime=-1}^j
    h_{\alpha\beta,i^\prime}^{(j^\prime)}
    I_{a\beta,i-i^\prime}^{(j-j^\prime,k)}
    -g_{\alpha,i}^{(j,k)}.
  \label{mi7}
\end{equation}
From (\ref{mi7}) together with (\ref{mi6.1}) we obtain the
subleading powers in $z$ of $I_{a1}$ and $I_{a2}$. 
We also obtain the coefficient of the $z^0\ln z$-term of $I_{a2}$,
which we already calculated in (\ref{mi6.1}). This means that the differential
equations provide a non-trivial check for method of regions, which was
used for the leading power calculation. 

\subsection{Diagrams of Fig. ~\ref{f1}b}
The topology Fig.~\ref{f1}b comes with the master integrals
\begin{eqnarray}
I_{b1} &=&
\intdtwo\frac{1}
{((k+p)^2-1)((k+p-q)^2-1)(l^2-z)((k+l)^2-z)}
\nonumber\\
I_{b2} &=&
\intdtwo\frac{1}
{((k+p)^2-1)((k+p-q)^2-1)(l^2-z)^2((k+l)^2-z)}
\nonumber\\
I_{b3} &=&
\intdtwo\frac{1}
{((k+p)^2-1)((k+p-q)^2-1)^2(l^2-z)((k+l)^2-z)}.
\label{mi8}
\end{eqnarray}
We need these integrals in leading power. Besides the hard region,
where all of these integrals get a leading power contribution,
$I_{b2}$ also gets contributions from two further regions. 
In the soft region defined by
$k^\mu\sim 1$ and $l^\mu\sim\sqrt{z}$ $I_{b2}$ becomes
\begin{equation}
I_{b2,s}=z^{-\epsilon}\intdtwo\frac{1}
{((k+p)^2-1)((k+p-q)^2-1)(l^2-1)^2k^2},
\label{mi9}
\end{equation}
which is a product of (\ref{i32m}) and a trivial tadpole integral.
In the collinear region defined by $k^\mu\sim 1, 
l^+\sim 1$, $l_\perp\sim\sqrt{z}$ and 
$l^-\sim z$,  $I_{b2}$ takes the form
\begin{equation}
I_{b2,c}=z^{-\epsilon}\intdtwo\frac{1}
{((k+p)^2-1)((k+p-q)^2-1)(l^2-1)^2(k^2+2l^+k^-)}.
\label{mi10}
\end{equation}
However, the collinear region has an overlap with the soft region,
where (\ref{mi10}) reduces to (\ref{mi9}). On the other hand
(\ref{mi10}) is indeed equal to (\ref{mi9}) which can be seen by the
following argument: Consider the integration $[dl]$. The integrand
depends besides on terms constant in $l^\mu$ only on $l^2$ and
$l^+=l\cdot n_-$. So $n^\mu_-$ is the only Lorentz vector that 
multiplies $l^\mu$. Because of Lorentz invariance the integral can
only depend on $n_-$ through $n_-^2=0$. So we can set $n_-$ to zero 
such that (\ref{mi10}) reduces to (\ref{mi9}). This is to say the collinear
region has already been taken into account by the soft region. To avoid
double counting we have to skip the contribution
(\ref{mi10}). Analogously we can introduce another collinear region
$k^\mu\sim 1, l\sim n_-$. 
By the same argument we see that also this region has been already taken
into account in (\ref{mi9}).

In the hard region IBP identities provide a reduction of (\ref{mi8})
to the set of integrals
\begin{eqnarray}
I_{b1,h} &=&
\intdtwo\frac{1}
{((k+p)^2-1)((k+p-q)^2-1)l^2(k+l)^2}
\nonumber\\
I_{b2,h} &=&
\intdtwo\frac{1}
{((k+p)^2-1)((k+p-q)^2-1)l^4(k+l)^2}.
\label{mi12}
\end{eqnarray}
These integrals can be evaluated via differential equations with
respect to $\hat{s}$. By defining 
\begin{equation}
  \vec{I}=\left({I_{b1,h}\atop I_{b2,h}}\right)
\end{equation}
and differentiating $\vec{I}$ with respect to $\hat{s}$ we obtain a
differential equation of the form
\begin{equation}
  \frac{d}{d\hat{s}}\vec{I}=h\vec{I}+\vec{g}
\label{mi13}
\end{equation}
where $\vec{g}$ contains the integrals (\ref{i3mb2z0}) and
(\ref{i3mb2z1}). We define the expansion in $\epsilon$
\begin{eqnarray}
  \vec{I}&=&\sum_{i=-2}^\infty \vec{I}^{(i)} \epsilon^i
  \nonumber\\
  h  &=& \sum_{i=0}^\infty h^{(i)} \epsilon^i
  \nonumber\\
  \vec{g}&=&\sum_{i=-2}^\infty \vec{g}^{(i)} \epsilon^i 
\end{eqnarray}
and write (\ref{mi13}) in the expanded form
\begin{eqnarray}
  \frac{d}{d\hat{s}}\vec{I}^{(-2)}&=&h^{(0)}\vec{I}^{(-2)}+\vec{g}^{(-2)}
  \nonumber\\
  \frac{d}{d\hat{s}}\vec{I}^{(-1)}&=&h^{(0)}\vec{I}^{(-1)}+h^{(1)}\vec{I}^{(-2)}+
  \vec{g}^{(-1)}
  \nonumber\\
  \frac{d}{d\hat{s}}\vec{I}^{(0)}&=&h^{(0)}\vec{I}^{(0)}+h^{(1)}\vec{I}^{(-1)}
  +h^{(2)}\vec{I}^{(-2)}
  +\vec{g}^{(0)}. 
\label{mi14} 
\end{eqnarray}
In our special case $h^{(0)}_{12}=0$ such that
(\ref{mi14}) decouples and we can solve (\ref{mi14}) by the common
methods \emph{separation of variables} and \emph{variation of
  constants}. From Feynman parameterization we see that the limit
$\hat{s}=0$
does not lead to additional divergences in $\epsilon$ and can be used as
initial condition for (\ref{mi14}). Finally we obtain
\begin{equation}
\begin{split}
&I_{b1,h} = \frac{1}{(4\pi)^4}\mu^{4\epsilon}
\Bigg[
\frac{-1}{2\epsilon^2}+
\frac{2
  \sqrt{\frac{4-\hat{s}}{\hat{s}}}\arcsin\frac{\sqrt{\hat{s}}}{2}
-\frac{5}{2}}
{\epsilon}+
\frac{3\left(\hat{s}-3\right)\arcsin^2\frac{\sqrt{\hat{s}}}{2}}{\hat{s}-1}
\\
&\quad
+\frac{-5\pi^2\hat{s}-114\hat{s}+7\pi^2+114}{12\left(\hat{s}-1\right)}
+\sqrt{\frac{4-\hat{s}}{\hat{s}}}
\Bigg(
\arcsin\frac{\sqrt{\hat{s}}}{2} \left(-2 \ln
    \left(4-\hat{s}\right)+
\ln\left(\hat{s}\right)+10\right)
\\
&\quad
+\text{Cl}_2\left(2\arcsin\frac{\sqrt{\hat{s}}}{2}\right)
-2\text{Cl}_2\left(2\arcsin\frac{\sqrt{\hat{s}}}{2}+\pi\right)
\Bigg)
\Bigg]
\\
&I_{b2,h} = \frac{1}{(4\pi)^4}\mu^{4\epsilon}
\Bigg[
\frac{6\arcsin^2\frac{\sqrt{\hat{s}}}{2}-\frac{\pi^2}{6}}
{\epsilon}+
\left(12\ln\left(1-\hat{s}\right)+
 3\ln\left(\hat{s}\right)\right) 
 \arcsin^2\frac{\sqrt{\hat{s}}}{2}
\\
&\quad
 +4\text{Cl}_2\left(6\arcsin\frac{\sqrt{\hat{s}}}{2}+\pi\right) 
  \arcsin\frac{\sqrt{\hat{s}}}{2}
  -\frac{\pi^2}{3}\ln\left(1-\hat{s}\right)
  -3 \text{Cl}_3\left(2\arcsin\frac{\sqrt{\hat{s}}}{2}\right)
\\
&\quad
  +6\text{Cl}_3\left(2 \arcsin\frac{\sqrt{\hat{s}}}{2}+\pi\right)
  +\frac{2}{3}\text{Cl}_3\left(6\arcsin\frac{\sqrt{\hat{s}}}{2}+\pi\right)
  +3\zeta (3)
\Bigg],
\end{split}
\label{mi15}
\end{equation}
where $\Cl(\phi)=\Im\Li(e^{i\phi})$ and 
$\Clthree(\phi)=\Re\Lithree(e^{i\phi})$.

\subsection{Diagrams of Fig. ~\ref{f1}c}
The topology Fig.~\ref{f1}c comes with the master integrals
\begin{eqnarray}
I_{c1} &=&
\intdtwo\frac{1}
{(l^2-z)((k+l)^2-z)((l+q)^2-z)(k+p-q)^2}
\nonumber\\
I_{c2} &=&
\intdtwo\frac{1}
{(l^2-z)^2((k+l)^2-z)((l+q)^2-z)(k+p-q)^2}
\nonumber\\
I_{c3} &=&
\intdtwo\frac{1}
{(l^2-z)((k+l)^2-z)^2((l+q)^2-z)(k+p-q)^2}.
\label{mi16}
\end{eqnarray}
They all get leading power contributions from the hard region, where
IBP identities lead to a further reduction of $I_{c2}$ and $I_{c3}$ to
$I_{c1}$. $I_{c1}$ can be calculated by a differential equation with
respect to $\hat{s}$, which reads:
\begin{equation}
\frac{d}{d\hat{s}}I_{c1,h}=
\epsilon\frac{2\hat{s}-1}{\hat{s}(1-\hat{s})}I_{c1,h}-
\kappa(\mu,\epsilon)\frac{1}{\hat{s}(1-\hat{s})},
\label{mi17}
\end{equation}
where
\begin{equation}
\kappa(\mu,\epsilon)=
\frac{\left(\mu^2e^{\gamma_E}\right)^{2\epsilon}}{(4\pi)^4}
e^{2i\pi\epsilon}
\frac{\Gamma(-1+2\epsilon)\Gamma^3(1-\epsilon)}{\Gamma(2-3\epsilon)}.
\end{equation}
The most general solution of (\ref{mi17}) is given by 
\begin{equation}
c\hat{s}^{-\epsilon}(1-\hat{s})^{-\epsilon}
-\kappa(\mu,\epsilon)
(1-\hat{s})^{-\epsilon}
\left[
\frac{{}_2\text{F}_1(-\epsilon,\epsilon;1+\epsilon;\hat{s})}{\epsilon}+
\frac{\hat{s}\,{}_2\text{F}_1(1-\epsilon,1+\epsilon;2+\epsilon;\hat{s})}
{1+\epsilon}
\right],
\label{mi18}
\end{equation}
where we have to determine $c$. We note that both $\hat{s}=0$ and
$\hat{s}=1$ are no appropriate initial conditions. Hence we determine
$c$ by calculating the term proportional to $\hat{s}^{-\epsilon}$ in 
the expansion
of $I_{c1,h}$ around $\hat{s}=0$. The Mellin-Barnes representation
(see e.g.\ \cite{smirnov:1990rz}) of $I_{c1,h}$ reads
\begin{equation}
\begin{split}
&I_{c1,h}=-\frac{\left(\mu^2e^{\gamma_E}\right)^{2\epsilon}}{(4\pi)^4}
e^{2i\pi\epsilon}
\int_{-i\infty}^{i\infty}dt\,\hat{s}^t\Gamma(-t)\Gamma(t+2\epsilon)
\int_0^1dx\,x^{-2\epsilon}(1-x)^t
\\
&\quad\times
\int_0^1d^2y\,y_1^{-1-\epsilon-t}(1-y_1)^{-\epsilon}
y_2^{-2\epsilon-t}(1-y_2)^{-\epsilon}(1-y_1y_2)^t.
\end{split}
\label{mi19}
\end{equation}
We have to calculate the residue at $t=-\epsilon$ in (\ref{mi19}), 
which arises due to the
integration $\int_0^1d^2y\,y_1^{-1-\epsilon-t}(\ldots)$ at $y_1=0$. So
we can set $y_1=0$ in the ellipsis and obtain
\begin{equation}
I_{c1,h}=\hat{s}^{-\epsilon}
\left[
-\frac{\left(\mu^2e^{\gamma_E}\right)^{2\epsilon}}{(4\pi)^4}
e^{2i\pi\epsilon}\frac{\Gamma^2(\epsilon)\Gamma^3(1-\epsilon)}
{(1-2\epsilon)\Gamma(2-3\epsilon)}
\right]+\ldots,
\label{mi20}
\end{equation}
where the ellipsis denotes integer powers of $\hat{s}$.
Hence $c$ reads
\begin{equation}
c=-\frac{\left(\mu^2e^{\gamma_E}\right)^{2\epsilon}}{(4\pi)^4}
e^{2i\pi\epsilon}\frac{\Gamma^2(\epsilon)\Gamma^3(1-\epsilon)}
{(1-2\epsilon)\Gamma(2-3\epsilon)}.
\label{mi21}
\end{equation}

In the collinear region $k^+,l^+\sim1$, $k_\perp,l_\perp\sim\sqrt{z}$,
$k^-,l^-\sim z$ both $I_{c2}$ and $I_{c3}$ get a leading power
contribution:
\begin{equation}
\begin{split}
&I_{c2,c}=z^{-2\epsilon}\intdtwo\frac{1}
{(l^2-1)^2((k+l)^2-1)(2l^+q^-+\hat{s})(k^2+2k^-(p-q)^+)}=
\\
&\quad
-\frac{\left(\mu^2e^{\gamma_E}\right)^{2\epsilon}}{(4\pi)^2}
\frac{\Gamma^2(\epsilon)}{2(1-\epsilon)}
\frac{z^{-2\epsilon}}{\hat{s}}
\,{}_3\text{F}_2\left(1,1,\epsilon;2-\epsilon,1+2\epsilon;
\frac{\hat{s}-1}{\hat{s}}\right)\\
&I_{c3,c}=z^{-2\epsilon}\intdtwo\frac{1}
{(l^2-1)((k+l)^2-1)^2(2l^+q^-+\hat{s})(k^2+2k^-(p-q)^+)}=
\\
&\quad
-\frac{\left(\mu^2e^{\gamma_E}\right)^{2\epsilon}}{(4\pi)^2}
\frac{\Gamma^2(\epsilon)}{2(1-\epsilon)}
\frac{z^{-2\epsilon}}{\hat{s}}
\,{}_3\text{F}_2\left(1,1,1+\epsilon;2-\epsilon,1+2\epsilon;
\frac{\hat{s}-1}{\hat{s}}\right),
\end{split}
\label{mi22}
\end{equation}
where ${}_3\text{F}_2$ is given by 
\begin{equation}
{}_3\text{F}_2(a_1,a_2,a_3;b_1,b_2;x)=
\sum_{n=0}^\infty\frac{\Gamma(a_1+n)\Gamma(a_2+n)\Gamma(a_3+n)}
{\Gamma(a_1)\Gamma(a_2)\Gamma(a_3)}
\frac{\Gamma(b_1)\Gamma(b_2)}{\Gamma(b_1+n)\Gamma(b_2+n)}
\frac{x^n}{n!},
\label{mi23}
\end{equation}
and can be expanded in $\epsilon$ by the tools developed in 
\cite{Huber:2005yg,Huber:2007dx}.

In the soft region $k^\mu+l^\mu\sim\sqrt{z}$ only $I_{c3}$ contributes in
leading power:
\begin{equation}
I_{c3,s}=z^{-\epsilon}
\intdtwo\frac{1}
{l^2(l+q)^2(l-p+q)^2(k^2-z)^2},
\label{mi24}
\end{equation}
which is a product of two simple one-loop integrals. There are two
further collinear regions $k+l\sim n_+$ and $k+l\sim n_-$, where
$I_{c3}$ obtains a leading power contribution. However by an argument
similar to that given in the previous subsection we can show that these
contributions have already been taken into account by (\ref{mi24}).

As described above the subleading powers of (\ref{mi16}) are obtained via
differential equations with respect to $z$. Like in the previous cases
the terms of the order $z^0\ln z$ provide a check that we have taken
all regions contributing at leading power consistently into account.

\subsection{Diagrams of Fig.~\ref{f1}d}
The topology in Fig.~\ref{f1}d comes with two sets of master
integrals.
\begin{eqnarray}
I_{d11}&=&
\intdtwo\frac{1}
{(l^2-z)((l-k)^2-z)((l-q)^2-z)((k-p)^2-1)}
\nonumber\\
I_{d12}&=&
\intdtwo\frac{1}
{(l^2-z)((l-k)^2-z)^2((l-q)^2-z)((k-p)^2-1)}
\nonumber\\
I_{d13}&=&
\intdtwo\frac{1}
{(l^2-z)((l-k)^2-z)((l-q)^2-z)^2((k-p)^2-1)}
\nonumber\\
I_{d14}&=&
\intdtwo\frac{1}
{(l^2-z)((l-k)^2-z)((l-q)^2-z)((k-p)^2-1)^2}
\label{mi25}
\end{eqnarray}
and
\begin{eqnarray}
I_{d21}&=&
\intdtwo\frac{1}
{k^2((l-k)^2-z)((l-q)^2-z)((k-p)^2-1)}
\nonumber\\
I_{d22}&=&
\intdtwo\frac{1}
{k^2((l-k)^2-z)((l-q)^2-z)^2((k-p)^2-1)}
\nonumber\\
I_{d23}&=&
\intdtwo\frac{1}
{k^2((l-k)^2-z)((l-q)^2-z)((k-p)^2-1)^2}.
\label{mi26}
\end{eqnarray}

Let us consider the first set (\ref{mi25}). In the hard region this
set reduces by IBP identities to $I_{d11}$ and $I_{d12}$. These
integrals can be calculated by differential equations with respect to
$\hat{s}$. We obtain a system of differential equations similar to
(\ref{mi14}) where we have to use $\hat{s}=1$ as initial condition
because the integrals diverge at $\hat{s}=0$. The matrix
$h^{(0)}$ has vanishing off-diagonal elements such that
the system of differential equations decouples. In addition the $h^{(i)}$
contain only terms of the form $1/(1-\hat{s})$, $1/\hat{s}$ and
$\hat{s}^n$. So we can reduce the integrals to harmonic
polylogarithms, which were defined in \cite{Remiddi:1999ew}. The way
to do this is very well described in Section 2.4 of
\cite{Bell:2007tz}. Finally we used the program described in
\cite{Maitre:2005uu,Maitre:2007kp} to convert harmonic polylogarithms
into common functions like polylogarithms.

The soft region $l^\mu-k^\mu\sim\sqrt{z}$ leads to a leading power
contribution of $I_{d12}$
\begin{equation}
I_{d12,s}=z^{-\epsilon}
\intdtwo\frac{1}{k^2(k-q)^2((k-p)^2-1)(l^2-1)^2},
\label{m27}
\end{equation}
where we substituted $l\to l+k$. This integral is a product of
(\ref{i31m}) and a simple one-loop tadpole integral. 

The soft region $l^\mu-q^\mu\sim\sqrt{z}$ leads to a leading power
contribution of $I_{d13}$
\begin{equation}
I_{d13,s}=\frac{z^{-\epsilon}}{\hat{s}}
\intdtwo\frac{1}{k^2(k^2-1)(l^2-1)^2},
\end{equation}
where we substituted $l\to l+q$. This integral is a product of two
simple one-loop integrals.

Let us consider the second set of master integrals (\ref{mi26}).
In the hard region the set reduces via IBP identities to $I_{d21}$.
We evaluated $I_{d21}$ by a differential equation with respect to
$\hat{s}$. Solving this differential equation is a straightforward
calculation, which is analogous to the way we solved (\ref{mi17}).

The soft region $l^\mu-q^\mu\sim\sqrt{z}$ leads to a leading power
contribution of $I_{d22}$
\begin{equation}
I_{d22,s}=z^{-\epsilon}
\intdtwo\frac{1}{k^2(k-q)^2((k-p)^2-1)(l^2-1)^2},
\end{equation}
which coincides with (\ref{m27}).

Besides the leading power we also need the order $z$ of
$I_{d21}$. It is straightforward to calculate the order $z$ contribution
of the hard region by expanding the integrand of $I_{d21}$ up to the order $z$.
Finally the soft regions $l^\mu-k^\mu\sim\sqrt{z}$ and  
$l^\mu-q^\mu\sim\sqrt{z}$ contribute at order $z$. 
Since these regions do not overlap we have to take both of them into
account. After an appropriate
shift of $l$, 
$I_{d21}$ can in both regions be cast into the form
\begin{equation}
I_{d21,s}=z^{1-\epsilon}\intdtwo\frac{1}
{k^2(k-q)^2((k-p)^2-1)(l^2-1)},
\end{equation}
which is similar to (\ref{m27}). 

\subsection{Diagrams of Fig.~\ref{f1}e and f}
The integrals occurring in the diagrams of Fig.~\ref{f1}e 
reduce to (\ref{i22m}), (\ref{i32z}), (\ref{i32z2}) and (\ref{i32z3}).
The topology of Fig.~\ref{f1}f factorizes trivially into two one-loop
topologies, which have already been evaluated exactly in $\hat{s}$ in
\cite{Asatryan:2001zw}. As already mentioned in 
Section \ref{s1}, Fig.~\ref{f1}f does not
contribute to the form factors $F_{1,2}^{(7,9)}$ by definition;
its effect is, however, absorbed 
into a modified Wilson coefficient $C_9$ as in \cite{Asatryan:2001zw}.

\section{Results \label{secres}}
\subsection{Results for the form factors $F_{1,2}^{(7,9)}$ in the high $q^2$ region}
\begin{figure}
\begin{tabular}{l@{\hspace{-0.1cm}}l}
\resizebox{0.52\textwidth}{!}
{
\input{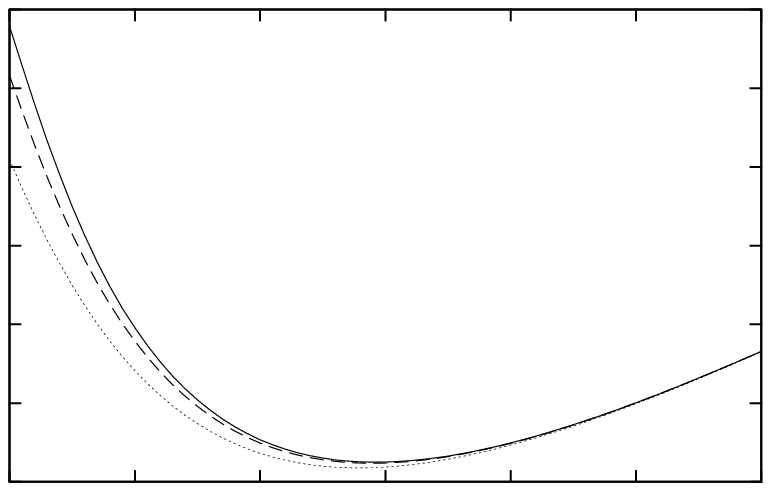}
}
&
\resizebox{0.52\textwidth}{!}
{
\input{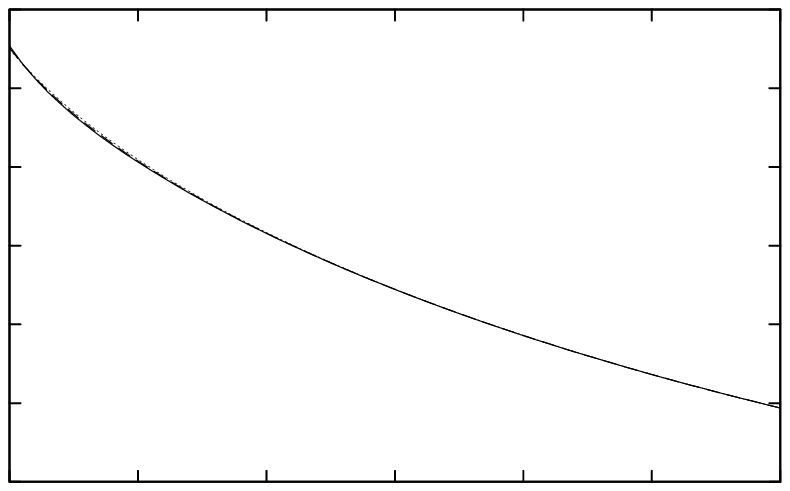}
}
\\[-0.1cm]
\resizebox{0.52\textwidth}{!}
{
\input{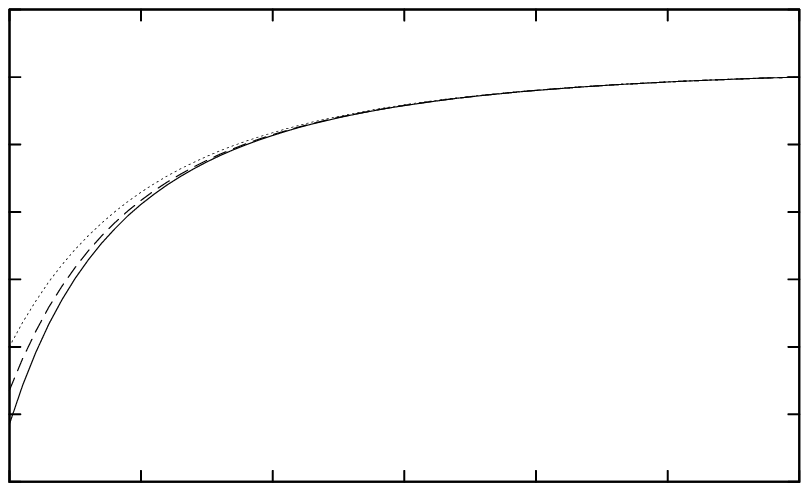}
}
&
\resizebox{0.52\textwidth}{!}
{
\input{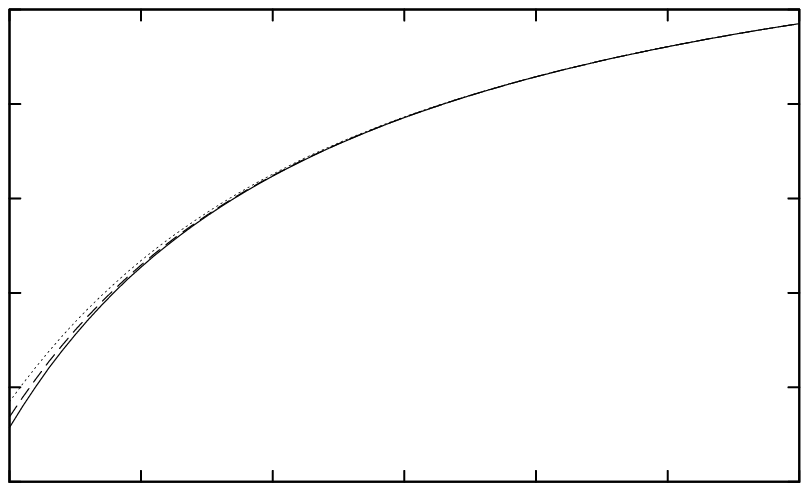}
}
\\[-0.1cm]
\resizebox{0.52\textwidth}{!}
{
\input{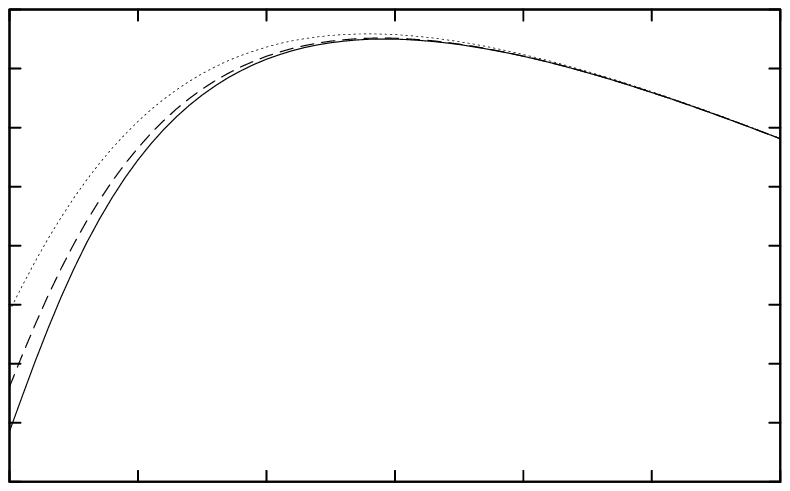}
}
&
\resizebox{0.52\textwidth}{!}
{
\input{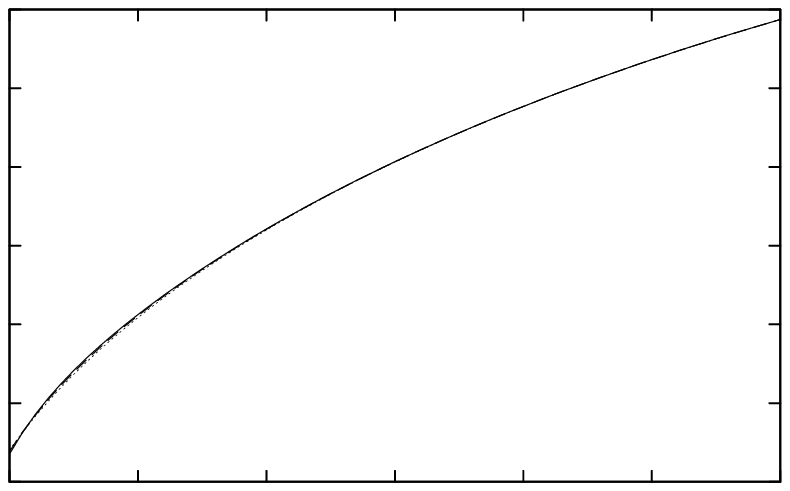}
}
\\[-0.1cm]
\resizebox{0.52\textwidth}{!}
{
\input{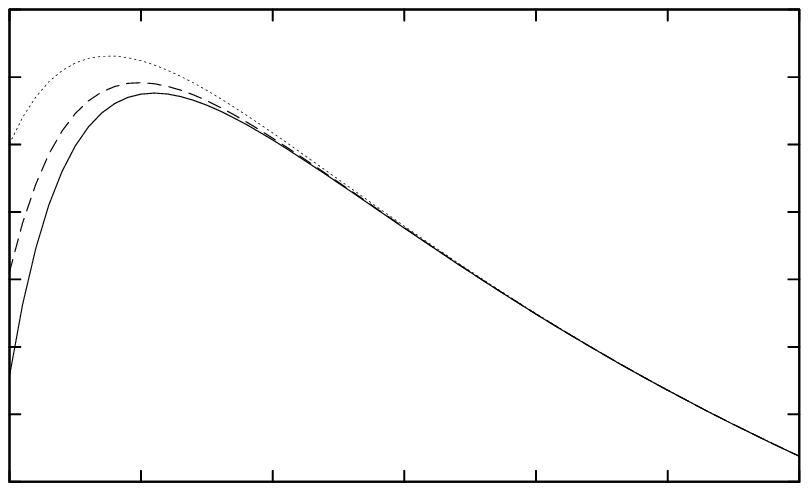}
}
&
\resizebox{0.52\textwidth}{!}
{
\input{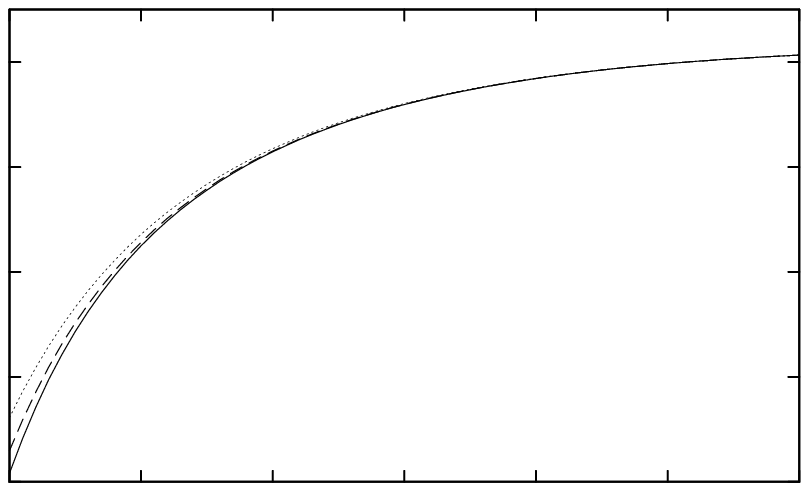}
}
\end{tabular}
\caption{Real and imaginary parts of the form factors
  $F_{1,2}^{(7,9)}$ as functions of
  $\hat{s}$. To demonstrate the convergence of the expansion in $z$ we
  included all orders up to $z^6$, $z^8$ and
  $z^{10}$ in the dotted, dashed and solid lines respectively. 
  We put $\mu=m_b$ and used the default value
  $z=0.1$.}

\label{f2}
\end{figure}
In Section \ref{s2} we calculated the two-loop diagrams
in Fig.~\ref{f1}a --e which contribute to the 
form factors $F_{1,2}^{(7,9)}$ defined in (\ref{formfact}). In addition,
there are counterterm contributions which have to be taken into
account. These counterterms are qualitatively the same as
those discussed in Section III.B of \cite{Asatryan:2001zw}.
Because its calculation in the high $q^2$-region is straightforward,
we do not list their explicit results. We only stress that in the
following results the $c$-quark mass is renormalized in the pole-scheme.

We calculated the renormalized form factors $F_{1,2}^{(7,9)}$ 
in the large $q^2$-region as  expansions of the form 
$c_{nm}(\hat{s}) z^n \ln^m z$ ($n=0,\frac{1}{2},1,\frac{3}{2},\ldots$; $m=0,1,2,\ldots$), keeping the full analytic dependence 
on $\hat{s}$ ($z=m_c^2/m_b^2$, $\hat{s}=q^2/m_b^2$). 
We included all orders up to $z^{10}$. 
To demonstrate the convergence of the power expansions,
we show in Fig.~\ref{f2} the form factors as functions of $\hat{s}$,
where we include all orders up to $z^6$, $z^8$ and $z^{10}$. We use as
default value $z = 0.1$ such that the $c\bar{c}$-threshold is located at
$\hat{s} = 0.4$. One sees from the figures that far away from the 
$c\bar{c}$-threshold, i.e.\ for
$\hat{s} > 0.6$, the expansions for all form factors are well behaved.

\begin{table}
\begin{tabular}{|c|c|c|c|}
\hline
$\sqrt{z}$ & $\hat{s}$ &$F_1^{(7)}$&$F_2^{(7)}$\\
\hline\hline
 & 0.6 & $-0.928-0.408i-0.856\ell$&$5.57+2.45i+5.14\ell$
\\\cline{2-4}
\raisebox{-2ex}[0ex][-2ex]{0.25} & 0.7 & $-0.909-0.458i-0.856\ell$&$5.45+2.75i+5.14\ell$
\\\cline{2-4}
 & 0.8 & $-0.888-0.500i-0.856\ell$&$5.33+3.00i+5.14\ell$
\\\cline{2-4}
 & 0.9 & $-0.867-0.535i-0.856\ell$&$5.20+3.21i+5.14\ell$
\\\hline\hline
 & 0.6 & $-0.919-0.347i-0.856\ell$&$5.52+2.08i+5.14\ell$
\\\cline{2-4}
\raisebox{-2ex}[0ex][-2ex]{0.27} & 0.7 & $-0.905-0.402i-0.856\ell$&$5.43+2.41i+5.14\ell$
\\\cline{2-4}
 & 0.8 & $-0.888-0.449i-0.856\ell$&$5.33+2.69i+5.14\ell$
\\\cline{2-4}
 & 0.9 & $-0.869-0.488i-0.856\ell$&$5.21+2.93i+5.14\ell$
\\\hline\hline
 & 0.6 & $-0.904-0.280i-0.856\ell$&$5.42+1.68i+5.14\ell$
\\\cline{2-4}
\raisebox{-2ex}[0ex][-2ex]{0.29} & 0.7 & $-0.896-0.342i-0.856\ell$&$5.38+2.05i+5.14\ell$
\\\cline{2-4}
 & 0.8 & $-0.883-0.393i-0.856\ell$&$5.30+2.36i+5.14\ell$
\\\cline{2-4}
 & 0.9 & $-0.867-0.437i-0.856\ell$&$5.20+2.62i+5.14\ell$
\\\hline\hline
 & 0.6 & $-0.879-0.208i-0.856\ell$&$5.28+1.25i+5.14\ell$
\\\cline{2-4}
\raisebox{-2ex}[0ex][-2ex]{0.31} & 0.7 & $-0.881-0.277i-0.856\ell$&$5.29+1.66i+5.14\ell$
\\\cline{2-4}
 & 0.8 & $-0.874-0.334i-0.856\ell$&$5.24+2.00i+5.14\ell$
\\\cline{2-4}
 & 0.9 & $-0.862-0.382i-0.856\ell$&$5.17+2.29i+5.14\ell$
\\\hline\hline
 & 0.6 & $-0.842-0.130i-0.856\ell$&$5.05+0.779i+5.14\ell$
\\\cline{2-4}
\raisebox{-2ex}[0ex][-2ex]{0.33} & 0.7 & $-0.858-0.207i-0.856\ell$&$5.15+1.24i+5.14\ell$
\\\cline{2-4}
 & 0.8 & $-0.859-0.269i-0.856\ell$&$5.15+1.62i+5.14\ell$
\\\cline{2-4}
 & 0.9 & $-0.853-0.322i-0.856\ell$&$5.12+1.93i+5.14\ell$
\\\hline
\end{tabular}

\caption{Numerical results for the form factors $F_{1,2}^{(7)}$, 
  for different values
  of $z$ and $\hat{s}$ ($\ell=\ln\frac{\mu}{m_b}$).}
\label{tab1}
\end{table}
\begin{table}
\begin{tabular}{|c|c|c|c|}
\hline
$\sqrt{z}$ & $\hat{s}$ &$F_1^{(9)}$&$F_2^{(9)}$\\
\hline\hline
 & 0.6 & $8.72-22.9i+(-5.47-3.23i)\ell-1.05\ell^2$&$13.2+13.6i+(22.1+19.4i)\ell+6.32\ell^2$
\\\cline{2-4}
\raisebox{-2ex}[0ex][-2ex]{0.25} & 0.7 & $9.53-19.8i+(-5.13-3.31i)\ell-1.05\ell^2$&$10.3+14.5i+(20.1+19.9i)\ell+6.32\ell^2$
\\\cline{2-4}
 & 0.8 & $9.92-17.5i+(-4.86-3.36i)\ell-1.05\ell^2$&$7.88+14.8i+(18.5+20.2i)\ell+6.32\ell^2$
\\\cline{2-4}
 & 0.9 & $10.1-15.8i+(-4.64-3.40i)\ell-1.05\ell^2$&$5.93+14.8i+(17.2+20.4i)\ell+6.32\ell^2$
\\\hline\hline
 & 0.6 & $7.65-26.6i+(-5.66-3.11i)\ell-1.05\ell^2$&$15.0+10.9i+(23.3+18.7i)\ell+6.32\ell^2$
\\\cline{2-4}
\raisebox{-2ex}[0ex][-2ex]{0.27} & 0.7 & $9.07-22.7i+(-5.29-3.23i)\ell-1.05\ell^2$&$12.0+12.6i+(21.1+19.4i)\ell+6.32\ell^2$
\\\cline{2-4}
 & 0.8 & $9.78-20.0i+(-5.00-3.30i)\ell-1.05\ell^2$&$9.44+13.3i+(19.3+19.8i)\ell+6.32\ell^2$
\\\cline{2-4}
 & 0.9 & $10.2-17.9i+(-4.76-3.35i)\ell-1.05\ell^2$&$7.35+13.6i+(17.9+20.1i)\ell+6.32\ell^2$
\\\hline\hline
 & 0.6 & $5.76-31.0i+(-5.88-2.95i)\ell-1.05\ell^2$&$16.6+7.46i+(24.6+17.7i)\ell+6.32\ell^2$
\\\cline{2-4}
\raisebox{-2ex}[0ex][-2ex]{0.29} & 0.7 & $8.11-26.2i+(-5.47-3.12i)\ell-1.05\ell^2$&$13.6+10.1i+(22.2+18.7i)\ell+6.32\ell^2$
\\\cline{2-4}
 & 0.8 & $9.32-22.8i+(-5.15-3.22i)\ell-1.05\ell^2$&$11.0+11.5i+(20.3+19.3i)\ell+6.32\ell^2$
\\\cline{2-4}
 & 0.9 & $9.98-20.3i+(-4.89-3.29i)\ell-1.05\ell^2$&$8.81+12.2i+(18.7+19.7i)\ell+6.32\ell^2$
\\\hline\hline
 & 0.6 & $2.65-35.9i+(-6.12-2.74i)\ell-1.05\ell^2$&$17.9+3.06i+(26.1+16.4i)\ell+6.32\ell^2$
\\\cline{2-4}
\raisebox{-2ex}[0ex][-2ex]{0.31} & 0.7 & $6.46-30.1i+(-5.67-2.98i)\ell-1.05\ell^2$&$15.1+7.05i+(23.4+17.9i)\ell+6.32\ell^2$
\\\cline{2-4}
 & 0.8 & $8.41-26.0i+(-5.32-3.12i)\ell-1.05\ell^2$&$12.5+9.24i+(21.3+18.7i)\ell+6.32\ell^2$
\\\cline{2-4}
 & 0.9 & $9.50-23.0i+(-5.04-3.21i)\ell-1.05\ell^2$&$10.3+10.5i+(19.6+19.3i)\ell+6.32\ell^2$
\\\hline\hline
 & 0.6 & $-2.28-41.7i+(-6.39-2.45i)\ell-1.05\ell^2$&$18.4-2.61i+(27.7+14.7i)\ell+6.32\ell^2$
\\\cline{2-4}
\raisebox{-2ex}[0ex][-2ex]{0.33} & 0.7 & $3.84-34.6i+(-5.89-2.79i)\ell-1.05\ell^2$&$16.3+3.20i+(24.7+16.8i)\ell+6.32\ell^2$
\\\cline{2-4}
 & 0.8 & $6.90-29.7i+(-5.51-2.99i)\ell-1.05\ell^2$&$13.9+6.42i+(22.4+17.9i)\ell+6.32\ell^2$
\\\cline{2-4}
 & 0.9 & $8.60-26.1i+(-5.20-3.12i)\ell-1.05\ell^2$&$11.7+8.31i+(20.5+18.7i)\ell+6.32\ell^2$
\\\hline
\end{tabular}

\caption{Numerical results for the form factors $F_{1,2}^{(9)}$, 
  for different values
  of $z$ and $\hat{s}$ ($\ell=\ln\frac{\mu}{m_b}$).}
\label{tab2}
\end{table}
In Tab.~\ref{tab1},\ref{tab2} we list numerical values of the form factors for
different values of $z$ and $\hat{s}$, retaining the dependence on the
renormalization scale $\mu$. 
We compared our values in
Tab.~\ref{tab1},\ref{tab2} with the numerical values
\cite{Hurth_priv} 
that were used in
\cite{Ghinculov:2003qd}. We obtain nearly
perfect agreement, i.e.\ the difference is always smaller than $1\%$. 

\begin{figure}
\input{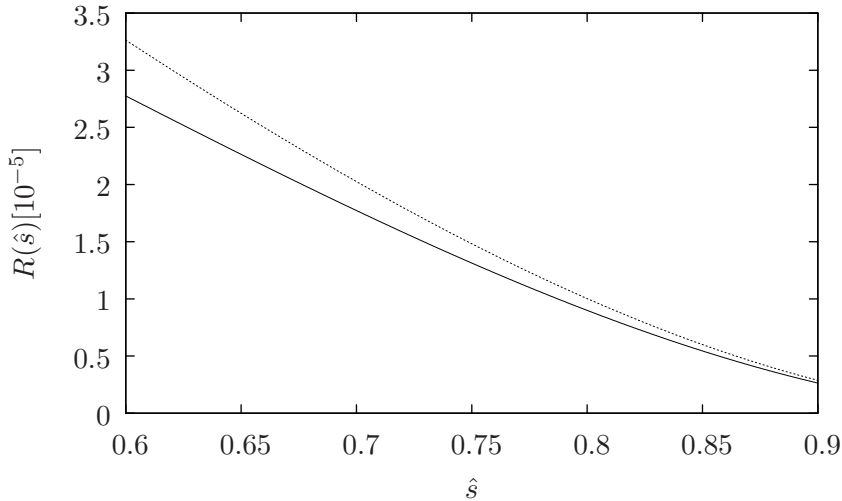}
\caption{Perturbative part of
$R(\hat{s})$ at NNLL. The solid line represents the NNLL
result, whereas in the dotted line the order $\alpha_s$ corrections 
to the matrix elements associated with $O_{1,2}$ are switched off.
We use $\mu=m_b$. See text for details.}
\label{f3}
\end{figure}

Unfortunately, the form factors are too lengthy to be given explicitly
in this paper.
Hence, the complete analytical results are attached to the source-code files
of the present paper at {\ttfamily www.arxiv.org}.
The Mathematica file \texttt{F\_high.m} contains the expressions for
\texttt{F17HighRe}, \texttt{F17HighIm},
\texttt{F19HighRe}, \texttt{F19HighIm}, \texttt{F27HighRe}, 
\texttt{F27HighIm}, \texttt{F29HighRe} and \texttt{F29HighIm}, 
which represent the real and
imaginary part of the form factors $F_{1,2}^{(7,9)}$ in the high
$\hat{s}$ region; they are defined in terms of \texttt{muh}, 
\texttt{z} and \texttt{sh} standing for $\mu/m_b$, $z$ and $\hat{s}$ 
respectively. Additionally this file contains the 
expressions for \texttt{DeltaF19HighRe},
\texttt{DeltaF19HighIm}, \texttt{DeltaF29HighRe} and
\texttt{DeltaF29HighIm}, which have to be added to the pole-scheme form
factors in order to switch from the pole-scheme to the
$\overline{\text{MS}}$-scheme of the $c$-quark mass.
For completeness we also provide 
the file \texttt{F\_low.m}, which contains the analogous
expressions in the low $\hat{s}$ region (\texttt{F17LowRe} etc.)
taken from \cite{Asatryan:2001zw}.
For numerical purposes we also provide the c++ header files 
\texttt{F\_1.h} and \texttt{F\_2.h} that contain the analogously 
defined functions\\[2ex]
{\ttfamily
\hspace*{1cm} double F\_17re(double muh, double z, double sh),\\
\hspace*{1cm} double F\_17im(double muh, double z, double sh),
} etc.\\[2ex]
valid in both high and low $\hat{s}$ region.
These files need for numerical evaluation of the harmonic
polylogarithms the header file \texttt{hpl.h}, which we provide at the
same place.

\subsection{Impact on the dilepton invariant mass spectrum in the high $q^2$ region}
In this section we briefly discuss 
the impact of the form factors $F_{1,2}^{(7,9)}$ calculated in this paper
on the $q^2$-spectrum at high values of $q^2$. 
To this end, we consider as in \cite{Asatryan:2001zw} 
the perturbative part of the ratio 
\begin{equation}
R(\hat{s})=\frac{1}{\Gamma(\bar{B}\to X_c e^- \bar{\nu}_e)}
\frac{d\Gamma(\bar{B}\to X_s\ell^+\ell^-)}{d\hat{s}},
\label{r1}
\end{equation}
where the formulas for the decay rates 
$\Gamma(b\to X_c e^- \bar{\nu}_e)$  
and $d\Gamma(b\to X_s\ell^+\ell^-)/d\hat{s}$
can be found e.g.\ in Section VI of \cite{Asatryan:2001zw}.
The parameterization of $d\Gamma(b\to X_s\ell^+\ell^-)/d\hat{s}$
as specified in (89) and (90) of \cite{Asatryan:2001zw}
is also valid in the high $q^2$ region. All the ingredients contained
in these two eqs. are available for arbitrary $q^2$, except
$F_{1,2,8}^{(7,9)}$. The expressions for $F_{1,2}^{(7,9)}$ were derived in 
the previous sections of this
paper in the high $q^2$ range. The calculations of the renormalized
form factors $F_{8}^{(7,9)}$ is
much easier and we therefore immediately give the results
(valid for arbitrary $\hat{s} \in [0,1]$):
\begin{eqnarray}
\label{eq:f87}
F_8^{(7)} = &&  \frac{4 \pi^2}{27} 
\frac{(2+\hat{s})}{(1-\hat{s})^4} - \frac{4}{9} 
\frac{(11-16 \hat{s}+8 \hat{s}^2)}{(1-\hat{s})^2} 
-\frac{8}{9} \frac{\sqrt{\hat{s}} \, \sqrt{4-\hat{s}}}{(1-\hat{s})^3}
(9 - 5 \hat{s}+2\hat{s}^2) \arcsin \left(\frac{\sqrt{\hat{s}}}{2} \right)
\nonumber \\
&& -\frac{16}{3} \frac{2+\hat{s}}{(1-\hat{s})^4} \arcsin^2 \left(
\frac{\sqrt{\hat{s}}}{2}  \right) - \frac{8 \hat{s}}{9(1-\hat{s})} \ln \hat{s}
-\frac{32}{9} \ln \frac{\mu}{m_b} -\frac{8}{9} \pi \, i
\end{eqnarray}
\begin{eqnarray}
\label{eq:f89}
F_8^{(9)} = && - \frac{8 \pi^2}{27} 
\frac{(4-\hat{s})}{(1-\hat{s})^4} + \frac{8}{9} 
\frac{(5-2 \hat{s})}{(1-\hat{s})^2} 
+\frac{16}{9} \frac{ \sqrt{4-\hat{s}}}{\sqrt{\hat{s}} \, (1-\hat{s})^3}
(4 +3 \hat{s}-\hat{s}^2) \arcsin \left( \frac{\sqrt{\hat{s}}}{2} \right)
\nonumber \\
&& +\frac{32}{3} \frac{(4-\hat{s})}{(1-\hat{s})^4} \arcsin^2 \left( 
\frac{\sqrt{\hat{s}}}{2} \right) + \frac{16}{9(1-\hat{s})} \ln \hat{s}
\end{eqnarray}
Fig.~\ref{f3} shows $R(\hat{s})$ defined in (\ref{r1}), 
where we set $\mu=5$ GeV and used
$\sqrt{z}=m_{c,\text{pole}}/m_{b,\text{pole}}=0.29$. We used the
pole-mass for the $b$-quark and the
$\overline{\text{MS}}$-mass for the top-quark and set
$m_{b,\text{pole}}=4.8$ GeV and 
$m_t(m_t)=163$ GeV \cite{Hoang:2008yj}.
We neglected the finite bremsstrahlung corrections calculated in 
\cite{Asatryan:2002iy}.
From Fig.~\ref{f3} we conclude that for $\mu=m_b$ the contributions of the
form factors $F_{1,2}^{(7,9)}$
lead to corrections of the order $10\% - 15\%$ at the level of the
perturbative part of the normalized $q^2$ spectrum $R(\hat{s})$.
\begin{figure}
\input{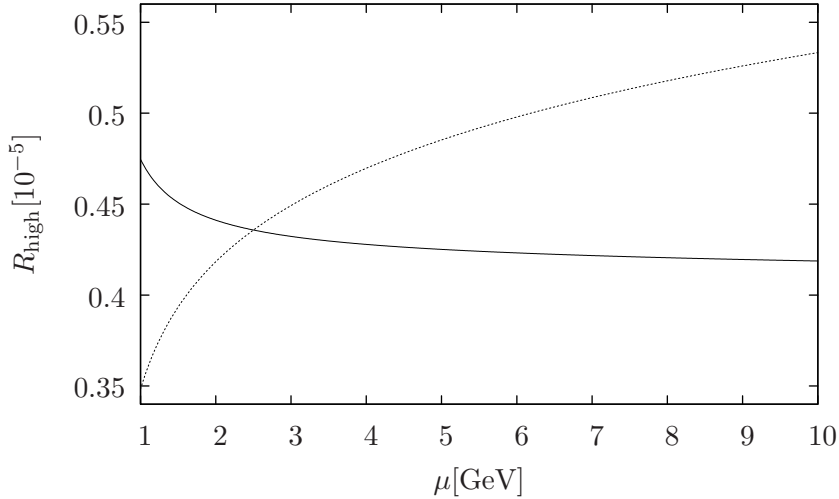}
\caption{Perturbative part of $R_\text{high}$ as function of 
the renormalization scale $\mu$ at 
NNLL. 
The solid line represents the NNLL
result, whereas in the dotted line the order $\alpha_s$ corrections 
to the matrix elements associated with $O_{1,2}$ are switched off.
See text for details.}
\label{f4}
\end{figure}
Integrating $R(\hat{s})$ over the high $\hat{s}$ region, we define
\begin{equation}
R_\text{high}=\int_{0.6}^1 d\hat{s}\, R(\hat{s}).
\end{equation}
Fig.~\ref{f4} shows the dependence of the perturbative part of
$R_\text{high}$ on the
renormalization scale. 
We obtain 
\begin{equation}
R_\text{high,pert}=(0.43\pm0.01(\mu))\times 10^{-5},
\end{equation}
where we determined the error by varying $\mu$ between 2 GeV and 
10 GeV. The
corrections due to the form factors $F_{1,2}^{(7,9)}$ 
lead to a decrease of the scale dependence to $2\%$.

We should mention at this point that a normalization different from the one in
(\ref{r1}) has been proposed in \cite{Ligeti:2007sn}: By normalizing the
$\bar{B} \to X_s \ell^+ \ell^-$ decay rate to the semileptonic
$\bar{B} \to X_u e^- \bar{\nu}_e$ decay rate with the same cut on $q^2$,
the large theoretical uncertainties due to power corrections can be 
significantly reduced. It was shown explicitly in a recent 
phenomenological update \cite{Huber:2007vv}
that the uncertainties from the poorly known
$O(1/m_b^3)$ power corrections are then under control. 
 
\section{Conclusions}
We calculated for the first time 
the NNLL virtual QCD corrections of the matrix elements of $\Op_1$ and $\Op_2$ 
in the high $q^2$ region as analytic functions of $q^2$ and $m_c$.
While keeping the full analytic dependence on $q^2$, we 
evaluated the matrix elements as an expansion in $z$ up to
the 10th power, which is numerically stable for $\hat{s}>0.6$.
Making extensive use of differential equation techniques, we
fully automatized the reduction of the higher order expansion
coefficients to the leading and first subleading power, which were
obtained via the method of regions.

Comparing our results for these matrix elements 
with those of a previous work where the master integrals were calculated
numerically \cite{Ghinculov:2003qd}, we obtain an agreement up to $1\%$.
Likewise in coincidence with \cite{Ghinculov:2003qd}, we find that the
corrections calculated in the present paper lead to a decrease of the
perturbative part of the
$q^2$-spectrum by $10\%-15\%$ relative to a NNLL result where these
contributions are not taken into account and reduce the renormalization
scale uncertainty to $2\%$.

We provide the rather lengthy results of our calculation in electronic
form as Mathematica files and for numerical purposes also as c++ files.

\section{Acknowledgments}
We would like to thank Thorsten Ewerth for initiating this project and for
collaboration at an early stage. We also thank H. Asatrian and U. Haisch for
helpful discussions.
This work is partially supported by the Swiss National Foundation as
well as EC-Contract MRTN-CT-2006-035482 (FLAVIAnet).
The Center for Research and Education in Fundamental Physics (Bern) is
supported by the ``Innovations- und Kooperationsprojekt C-13 of
the Schweizerische Universit\"atskonferenz SUK/CRUS''.

\newpage

\begin{appendix}
\section{Common master integrals\label{cm}}
All integrals are evaluated in $d=4-2\epsilon$ dimensions. 
In the following notation we suppress the positive imaginary part
$+i0$ of the denominators. 
The integration measure is defined as
\begin{equation}
[dk]=\left(\frac{\mu^2}{4\pi}e^{\gamma_E}\right)^\epsilon\frac{d^dk}{(2\pi)^d}.
\end{equation}

\subsection{One-loop integrals}
\subsubsection{2-point integral with two massive lines}
\begin{equation}
\raisebox{-.45cm}{\resizebox{2cm}{!}{\includegraphics{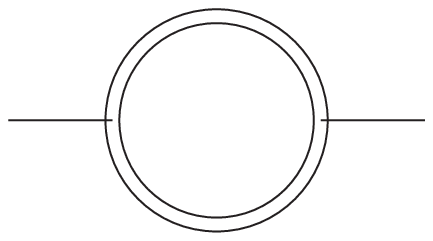}}} =
\intd \frac{1}{(k^2-m^2)((k+q)^2-m^2)}
\label{i22m}
\end{equation}
The double line denotes the massive propagator. We evaluate
(\ref{i22m}) in the two regions $q^2<4m^2$ and $q^2>4m^2$.
In the latter one we need the integral in an expansion 
in $m^2/q^2$. Using Mellin-Barnes
representation \cite{greub:1996tg,smirnov:2002pj} it is easily seen
that we can cast (\ref{i22m}) into the following form:
\begin{equation}
\frac{i}{(4\pi)^2}\left(\frac{\mu^2 e^{\gamma_E}}{q^2}\right)^\epsilon
\frac{1}{2\pi i}
\int_{-i\infty}^{i\infty}dt\,
\left(\frac{m^2}{q^2}\right)^t e^{i\pi(t+\epsilon)}
\frac{\Gamma(-t)\Gamma(t+\epsilon)\Gamma^2(1-t-\epsilon)}
{\Gamma(2-2t-2\epsilon)},
\end{equation}
where the integration contour over $t$ has to be chosen such that
$-\epsilon<\Re(t)<0$. The poles on the right hand side of the contour
are located at $t=n$ and $t=n+1-\epsilon$ where $n\in \mathbb{N}_0$. 
By closing the integration contour to the right
we obtain the power expansion in $m^2/q^2$.

Now let us consider the region $q^2<4m^2$. Up to order $\epsilon$ the
integral reads:
\begin{equation}
\begin{split}
&\raisebox{-.45cm}{\resizebox{2cm}{!}{\includegraphics{figs/I22m}}} =
\frac{i}{(4\pi)^2}\left(\frac{\mu^2}{m^2}\right)^\epsilon
\times
\\
&\quad \Bigg[\frac{1}{\epsilon}+
\sqrt{\frac{4-\hat{x}}{\hat{x}}}\arcsin\frac{\sqrt{\hat{x}}}{2}+
\\
&\quad
\epsilon\left(
4+\frac{\pi^2}{12}+
\sqrt{\frac{4-\hat{x}}{\hat{x}}}\arcsin\frac{\sqrt{\hat{x}}}{2}
\left(-4+2\ln(4-\hat{x})\right)+
2\sqrt{\frac{4-\hat{x}}{\hat{x}}}
\Cl\left(2\arcsin\frac{\sqrt{\hat{x}}}{2}+\pi\right)
\right)\bigg],
\end{split}
\end{equation}
where we defined $\hat{x}=q^2/m^2$ and $\Cl(\phi)=\Im\Li(e^{i\phi})$.

\subsubsection{3-point integral with one massive line}
\begin{equation}
\raisebox{-.45cm}{\resizebox{2cm}{!}{\includegraphics{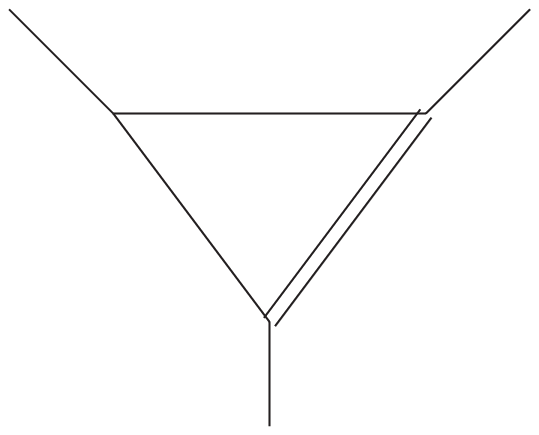}}} =
\intd \frac{1}{k^2(k+q)^2((k+p)^2-m^2)},
\label{i31m}
\end{equation}
where 
\begin{equation}
p^2=m^2,\quad q^2=\hat{x}m^2\quad\text{and}\quad\hat{x}<1.
\end{equation}
The integral is evaluated to
\begin{equation}
\begin{split}
&\quad\raisebox{-.45cm}{\resizebox{2cm}{!}{\includegraphics{figs/I31m}}} =
\frac{i}{(4\pi)^2}\left(\frac{\mu^2e^{\gamma_E}}{m^2}\right)^\epsilon
\frac{1}{m^2}\frac{\Gamma(\epsilon)}{\Gamma(2-2\epsilon)}\\
&\quad
\times\left[
\Gamma(1-2\epsilon)\,{}_2\text{F}_1(1,1;2-2\epsilon;1-\hat{x})-
\hat{x}^{-\epsilon}e^{i\pi\epsilon}
\Gamma^2(1-\epsilon)\,
{}_2\text{F}_1(1,1-\epsilon;2-2\epsilon;1-\hat{x})
\right],
\end{split}
\end{equation}
with ${}_2\text{F}_1$ given by (\ref{mi4.1}).
\subsubsection{3-point integral with two massive lines}
\begin{equation}
\raisebox{-.45cm}{\resizebox{2cm}{!}{\includegraphics{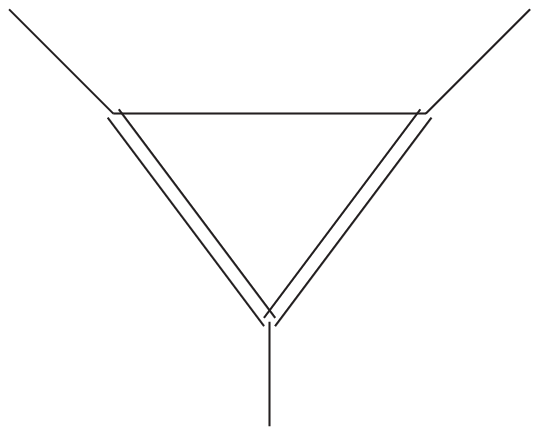}}} =
\intd \frac{1}{k^2((k+p)^2-m^2)((k+p-q)^2-m^2)},
\label{i32m}
\end{equation}
where
\begin{equation}
p^2=m^2,\quad q^2=\hat{x}m^2,\quad (p-q)^2=0\quad\and\quad
q^2<4m^2.
\end{equation}
The expansion in $\epsilon$ of (\ref{i32m}) reads
\begin{equation}
\begin{split}
&\raisebox{-.45cm}{\resizebox{2cm}{!}{\includegraphics{figs/I32m}}}=
\frac{i}{(4\pi)^2}\left(\frac{\mu^2}{m^2}\right)^\epsilon
\frac{1}{m^2(1-\hat{x})}
\times
\\
&\quad
\Bigg[-\frac{\pi^2}{6}+6\arcsin^2\frac{\sqrt{\hat{x}}}{2}+\\
&\quad \epsilon\bigg(
12 \ln\left(1-\hat{x}\right) \arcsin^2\frac{\sqrt{\hat{x}}}{2}+
4\text{Cl}_2\left(6\arcsin\frac{\sqrt{\hat{x}}}{2}+\pi\right) 
\arcsin\frac{\sqrt{\hat{x}}}{2}-\\
&\quad
\frac{1}{3}\pi ^2\ln\left(1-\hat{x}\right)+
6\text{Cl}_3\left(2\arcsin\frac{\sqrt{\hat{x}}}{2}+\pi\right)+
\frac{2}{3} 
\text{Cl}_3\left(6\arcsin\frac{\sqrt{\hat{x}}}{2}+\pi\right)+
2\zeta(3)
\bigg)
\Bigg],
\end{split}
\end{equation}
where $\Clthree(\phi)=\Re\Lithree(e^{i\phi})$

\subsection{Two-loop integrals}
\subsubsection{Two massive lines}
We need the following three sunrise diagrams in an expansion in
$m^2/q^2$. So as above we give the Mellin-Barnes representation, from
where the expansion can be easily derived.

\begin{equation}
\begin{split}
&\raisebox{-.45cm}{\resizebox{2cm}{!}{\includegraphics{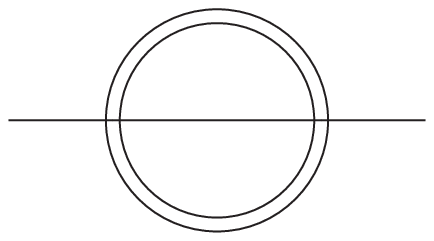}}}=
\intdtwo\frac{1}{(k+q)^2(l^2-m^2)((k+l)^2-m^2)}=
\\
&\quad
-\frac{1}{(4\pi)^4}q^2\left(\frac{\mu^2e^{2\gamma_E}}{q^2}\right)^\epsilon
\Gamma(1-\epsilon)
\frac{1}{2\pi i}\int_{-i\infty}^{i\infty}dt\,
\left(\frac{m^2}{q^2}\right)^te^{i\pi(2\epsilon+t)}\times
\\
&\quad
\frac{\Gamma(-t)\Gamma(t-1+2\epsilon)
\Gamma^2(1-\epsilon-t)\Gamma(2-2\epsilon-t)}
{\Gamma(2-2\epsilon-2t)\Gamma(3-3\epsilon-t)}.
\end{split}
\label{i32z}
\end{equation}
The residues we have to take into account are located at $n$, 
$n+1-\epsilon$ and
$n+2-2\epsilon$
with $n\in\mathbb{N}_0$.

\begin{equation}
\begin{split}
&\raisebox{-.45cm}{\resizebox{2cm}{!}{\includegraphics{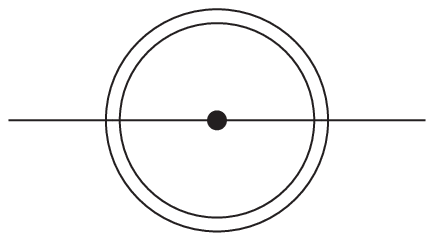}}}=
\intdtwo\frac{1}{\left[(k+q)^2\right]^2(l^2-m^2)((k+l)^2-m^2)}=
\\
&\quad
-\frac{1}{(4\pi)^4}\left(\frac{\mu^2e^{2\gamma_E}}{q^2}\right)^\epsilon
\Gamma(-\epsilon)
\frac{1}{2\pi i}\int_{-i\infty}^{i\infty}dt\,
\left(\frac{m^2}{q^2}\right)^te^{i\pi(2\epsilon+t)}\times
\\
&\quad
\frac{\Gamma(-t)\Gamma(t+2\epsilon)
\Gamma^2(1-\epsilon-t)\Gamma(2-2\epsilon-t)}
{\Gamma(2-2\epsilon-2t)\Gamma(2-3\epsilon-t)}.
\end{split}
\label{i32z2}
\end{equation}
The dotted line denotes a propagator that has to be taken squared.
The residues are located at $n$, $n+1-\epsilon$,
$n+2-2\epsilon$, $n\in\mathbb{N}_0$.

\begin{equation}
\begin{split}
&\raisebox{-.45cm}{\resizebox{2cm}{!}{\includegraphics{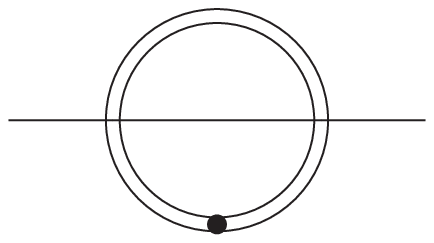}}}=
\intdtwo\frac{1}{(k+q)^2(l^2-m^2)^2((k+l)^2-m^2)}=
\\
&\quad
-\frac{1}{(4\pi)^4}\left(\frac{\mu^2e^{2\gamma_E}}{q^2}\right)^\epsilon
\Gamma(1-\epsilon)
\frac{1}{2\pi i}\int_{-i\infty}^{i\infty}dt\,
\left(\frac{m^2}{q^2}\right)^te^{i\pi(2\epsilon+t)}\times
\\
&\quad
\frac{\Gamma(-t)\Gamma(t+2\epsilon)
\Gamma(-\epsilon-t)\Gamma(1-\epsilon-t)\Gamma(1-2\epsilon-t)}
{\Gamma(1-2\epsilon-2t)\Gamma(2-3\epsilon-t)},
\end{split}
\label{i32z3}
\end{equation}
with the residues located at $n$, $n-\epsilon$, $n+1-\epsilon$, 
$n\in\mathbb{N}_0$.

\subsubsection{Three massive lines}
We need the following three integrals in an expansion in
$m^2/M^2$. Therefore we give their Mellin-Barnes representation.

\begin{equation}
\begin{split}
&\raisebox{-.45cm}{\resizebox{2cm}{!}{\includegraphics{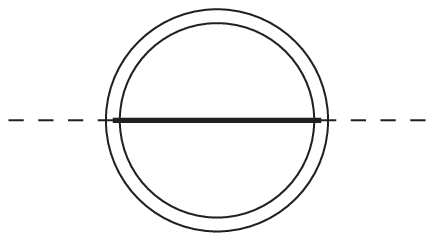}}}=
\intdtwo\frac{1}{(k^2-M^2)(l^2-m^2)^2((k+l)^2-m^2)}=
\\
&\quad
\frac{1}{(4\pi)^4}M^2\left(\frac{\mu^2e^{2\gamma_E}}{M^2}\right)^\epsilon
\frac{1}{2\pi i}\int_{-i\infty}^{i\infty}dt\,
\left(\frac{m^2}{M^2}\right)^t\times
\\
&\quad
\frac{\Gamma(-t)\Gamma(t-1+2\epsilon)
\Gamma^2(1-\epsilon-t)\Gamma(\epsilon+t)\Gamma(2-2\epsilon-t)}
{\Gamma(2-2\epsilon-2t)\Gamma(2-\epsilon)},
\end{split}
\label{i3mb2z0}
\end{equation}
with the residues located at $n$, $n+1-\epsilon$, $n+2-2\epsilon$,
$n\in\mathbb{N}_0$.

\begin{equation}
\begin{split}
&\raisebox{-.45cm}{\resizebox{2cm}{!}{\includegraphics{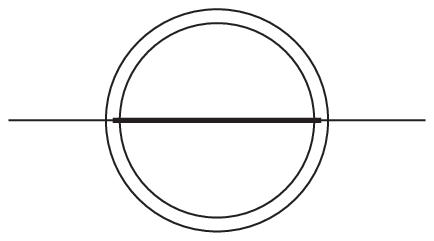}}}=
\intdtwo\frac{1}{((k+p)^2-M^2)(l^2-m^2)((k+l)^2-m^2)}=
\\
&\quad
\frac{1}{(4\pi)^4}M^2\left(\frac{\mu^2e^{2\gamma_E}}{M^2}\right)^\epsilon
\frac{1}{2\pi i}\int_{-i\infty}^{i\infty}dt\,
\left(\frac{m^2}{M^2}\right)^t\times
\\
&\quad
\frac{\Gamma(-t)\Gamma(t-1+2\epsilon)
\Gamma^2(1-\epsilon-t)\Gamma(\epsilon+t)\Gamma(3-4\epsilon-2t)}
{\Gamma(2-2\epsilon-2t)\Gamma(3-3\epsilon-t)},
\end{split}
\label{i3mb2z1}
\end{equation}
with $p^2=M^2$ and the residues located at $n$, $n+1-\epsilon$, 
$n/2+3/2-2\epsilon$, $n\in\mathbb{N}_0$.

\begin{equation}
\begin{split}
&\raisebox{-.45cm}{\resizebox{2cm}{!}{\includegraphics{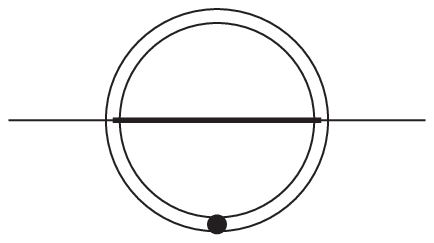}}}=
\intdtwo\frac{1}{((k+p)^2-M^2)(l^2-m^2)^2((k+l)^2-m^2)}=
\\
&\quad
-\frac{1}{(4\pi)^4}\left(\frac{\mu^2e^{2\gamma_E}}{M^2}\right)^\epsilon
\frac{1}{2\pi i}\int_{-i\infty}^{i\infty}dt\,
\left(\frac{m^2}{M^2}\right)^t\times
\\
&\quad
\frac{\Gamma(-t)\Gamma(t+2\epsilon)
\Gamma(-\epsilon-t)\Gamma(1-\epsilon-t)
\Gamma(1+\epsilon+t)\Gamma(1-4\epsilon-2t)}
{\Gamma(1-2\epsilon-2t)\Gamma(2-3\epsilon-t)},
\end{split}
\label{i3mb2z2}
\end{equation}
with $p^2=M^2$ and the residues located at $n$, $n+1-\epsilon$, 
$n/2+1/2-2\epsilon$, $n\in\mathbb{N}_0$.

\end{appendix}

\end{document}